\documentclass[aip, amsmath,amssymb,reprint]{revtex4-1}

\usepackage{graphicx,color,upgreek}
\usepackage{mathtools}
\usepackage{amsmath}

\let\oldAA\AA
\renewcommand{\AA}{\text{\normalfont\oldAA}}
\usepackage{textcomp}
\usepackage{color}
\usepackage{amssymb,amsmath} 
\usepackage{siunitx}

\newcommand{\rc}{r_\mathrm{cut}}

\newcommand{\paddyspeaks}[1]{{\color{black} #1}}

\begin{document}

\title{Direct Imaging of Contacts and Forces in Colloidal Gels}

\author{Jun Dong}
\affiliation{H.H. Wills Physics Laboratory, Tyndall Ave., Bristol, BS8 1TL, UK}
\affiliation{Centre for Nanoscience and Quantum Information, Tyndall Avenue, Bristol BS8 1FD, UK}
\affiliation{Bayer AG, Alfred Nobel Str. 50, 40789 Monheim, Germany}

\author{Francesco Turci}
\affiliation{H.H. Wills Physics Laboratory, Tyndall Ave., Bristol, BS8 1TL, UK}
\affiliation{Centre for Nanoscience and Quantum Information, Tyndall Avenue, Bristol BS8 1FD, UK}

\author{Robert L. Jack}
\affiliation{Department of Chemistry, University of Cambridge, Lensfield Road, Cambridge CB2 1EW, United Kingdom}
\affiliation{Department of Applied Mathematics and Theoretical Physics, University of Cambridge, Wilberforce Road, Cambridge CB3 0WA, United Kingdom}

\author{Malcolm A. Faers}
\affiliation{Bayer AG, Alfred Nobel Str. 50, 40789 Monheim, Germany}

\author{C. Patrick Royall}
\affiliation{Gulliver UMR CNRS 7083, ESPCI Paris, Universit\' e PSL, 75005 Paris, France.}
\affiliation{School of Chemistry, Cantock's Close, University of Bristol, BS8 1TS, UK}
\affiliation{Centre for Nanoscience and Quantum Information, Tyndall Avenue, Bristol BS8 1FD, UK}
\affiliation{H.H. Wills Physics Laboratory, Tyndall Ave., Bristol, BS8 1TL, UK}

\begin{abstract}
Colloidal dispersions are prized as model systems to understand basic properties of materials, and are central to a wide range of industries from cosmetics to foods to agrichemicals. Among the key developments in using colloids to address challenges in condensed matter is to resolve the particle coordinates in 3D, allowing a level of analysis usually only possible in computer simulation. However in amorphous materials, relating mechanical properties to microscopic structure remains problematic. This makes it rather hard to understand, for example, mechanical failure. Here we address this challenge by studying the \emph{contacts} and the \emph{forces} between particles, as well as their positions. To do so, we use a colloidal model system (an emulsion) in which the interparticle forces and local stress can be linked to the microscopic structure. We demonstrate the potential of our method to reveal insights into the failure mechanisms of soft amorphous solids by determining local stress in a colloidal gel. In particular, we identify ``force chains'' of load--bearing droplets, and local stress anisotropy, and investigate their connection with locally rigid packings of the droplets.
\end{abstract}

\maketitle

\section{Introduction}

A longstanding aim for studies of soft solids is to understand the mechanisms by which they fail or yield, either due to internal stresses or to imposed shear or other external fields \cite{cipelletti2005,fielding2014,divoux2016,bonn2017,nicolas2018,datta2011}. Theoretical approaches can be limited since these materials are often far--from--equilibrium and their properties depend on the details of the preparation protocol and mechanical history. This is problematic because yielding processes are often heterogeneous and in tackling this challenge it is useful to think of localised irreversible (or plastic) rearrangement events driven by stress at the microscopic scale \cite{bonn2017,vandervaart2013,nicolas2018}. On the micro-scale, the stress is a fluctuating quantity that is intrinsically linked to the packing of the particles i.e. the local structure. However, on passing to the macro-scale, the fluctuations in stress are no longer apparent, and the system obeys a constitutive relation that relates applied forces (stress) and material response (strain).

An important class of soft solids is colloidal gels \cite{royall2021}, which are encountered in numerous foods \cite{ubbink2012}, cosmetics, coatings, crop protection suspensions and pharmaceutical formulations. In addition to colloidal systems, a wide range of materials also exhibit gelation including proteins  \cite{mcmanus2016,cheng2021} phase-demixing oxides \cite{bouttes2014}, and metallic glassformers \cite{baumer2013}. The spatial inhomogeneity in colloidal gels means that gravitational stresses can become important, leading the system to collapse under its own weight \cite{harich2016}.  This last effect is an important determinant of the shelf-life of industrial products such as agri-chemicals. Among the most challenging aspects of gel collapse is that prior to  collapse, the elastic modulus of the gel \emph{increases}, so it becomes harder before it fails \cite{bartlett2012}.

A promising way to address phenomena such as gel failure is to use particle--resolved studies where the coordinates of individual particles are tracked \cite{royall2021,hunter2012}. In soft amorphous solids such as colloidal glasses, this technique has been used to image local re--arrangements of colloidal particles which may 
be precursors to large--scale failure \cite{schall2007,besseling2007}. In colloidal gels, particle resolved studies have revealed the rich nature of their local structure \cite{dinsmore2002,gao2007,royall2015prl,dibble2006,whitaker2019}. So far, while investigations of gel failure have related yielding to local crystallization  \cite{smith2007} and ingenious combinations of rheological methods and simulation and scattering have revealed the role of local plasticity \cite{vandoorn2017} and bulk two--point structure \cite{aime2018}, direct imaging of particle rearrangements have largely focussed on colloidal glasses  \cite{schall2007,besseling2007,royall2021} rather than gels \cite{lindstrom2012}.

However, rather than imaging of particle coordinates, an alternative route to understanding gel failure is to consider the local stress, as one expects that regions of high stress are where failure may occur. Now the local stress is manifested in the \emph{forces} between the particles. While using particle--resolved studies to obtain the coordinates of the particles is useful \cite{royall2021,besseling2007,schall2007}, it is clear that a major development would be some means to determine the force that each particle is under. This is in principle possible from measurement of the coordinates and knowledge of the interaction potential between the particles. While the latter can be estimated to a good approximation \cite{ivlev,royall2021}, the inevitable errors in determination of particle positions and polydispersity of the particles mean that it is very hard to convert the distance between two particles into a potential energy or force. This  means that this kind of measurement has hitherto only been possible in very special circumstances where the force varies slowly as a function of particle separation and the particles are far apart such that their positions, relative to the lengthscale over which the force varies can be very accurately determined \cite{assoud2009}. Thus from coordinate data only, stress correlations are typically inferred \emph{indirectly} \cite{hassani2018}. Computer simulations of course also provide access to coordinate data and to the forces between particles \cite{bouzid2017}. Often similar data is obtained
to that of particle--resolved experiments \cite{royall2018jcp}, particularly if hydrodynamic interactions are included  \cite{furukawa2010,royall2015prl,degraaf2019}. However for phenomena pertinent to failure in colloidal gels \cite{aime2018} and particularly delayed collapse \cite{bartlett2012}, the timescales and system sizes lie beyond those accessible to particulate simulation.

In granular systems, forces have been characterized for particles with diameters of at least $10\mu$m~\cite{brujic2003,brujic2007,desmond2013,jorjadze2011,jorjadze2013} (and often cm~\cite{majmudar2005,brodu2015}) and potential has been demonstrated for a scaled up version of a popular colloidal model system \cite{suhina2015}. Unlike athermal granular systems, the thermal motion exhibited by colloids leads to a multitude of new phenomena, such as the emergence of long--lived networks in the colloidal gels that we are interested in here. Identifying contacts and forces in colloidal systems is challenging due to sub-resolution length scales relevant to obtain forces between colloids. Here we take a first step to address this challenge, by investigating interparticle contacts and forces in colloidal gels via high--resolution optical microscopy. We use an emulsion system with a \emph{solvatochromic} dye, which is sensitive to the compressive forces between droplets ~\cite{brujic2007}. In this way, we obtain force contacts between droplets, and measures of the local stress. These we correlate via structural quantities and compare with computer simulation. We find that droplets in local structures associated with rigidity are more likely to be under higher pressure.

This paper is organized as follows. In section \ref{sectionMethods}, we explain our experimental protocol to identify contacts between particles, and proceed to describe how we may determine a local measure of compressive stress and how these are connected to form force chains. We detail the computer simulation methodology that we use to validate our experimental results. In our results section \ref{sectionResults} we compare the experimental results for the number of contacts per droplet and their coordination, with simulation. We go on to consider the distribution of compressive forces. We then investigate the length of force chains and compare these with computer simulation. Finally, we consider correlations between some of the quantities we have investigated. We discuss our findings in section \ref{sectionDiscussion}.

\begin{figure*}[hbt]
\centering
\includegraphics[width=160mm]{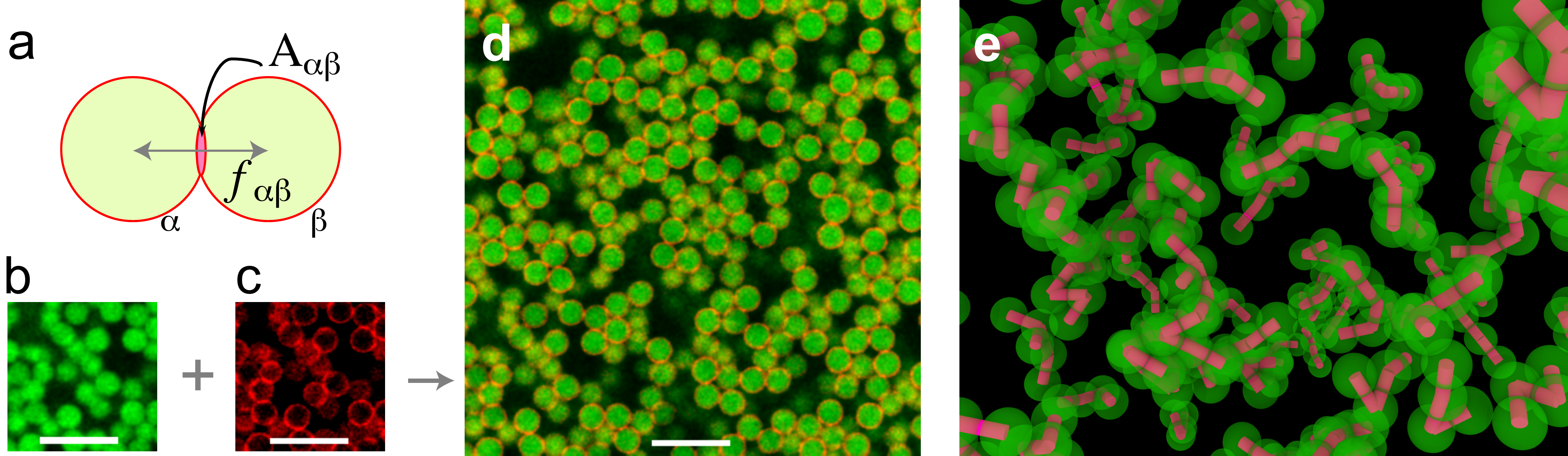}
\caption{\textbf{Visualising contacts and forces in emulsion gels.} 
(a) Schematic of distribution and fluorescence of solvatochromic dye two droplets $\alpha$ and $\beta$ at contact in an emulsion gel. The region shaded in pink $A_{\alpha \beta}$ is related to the force between droplets $f_{\alpha \beta}$. 
(b,c) Separate channels showing droplets (green) and contacts (red).
(d) Combined two--channel image of gel with $\phi_c=0.29$ and $c_p/c_p^\mathrm{gel}=1.50$.
(e) Contacts and particle coordinates identified in a gel. Rendered coordinates with compressive contacts indicated as pink sticks.}
\label{figContacts} 
\end{figure*}

\vspace{10mm}
\section{Methods}
\label{sectionMethods}

\begin{figure}[b]
\centering
\includegraphics[width=55mm]{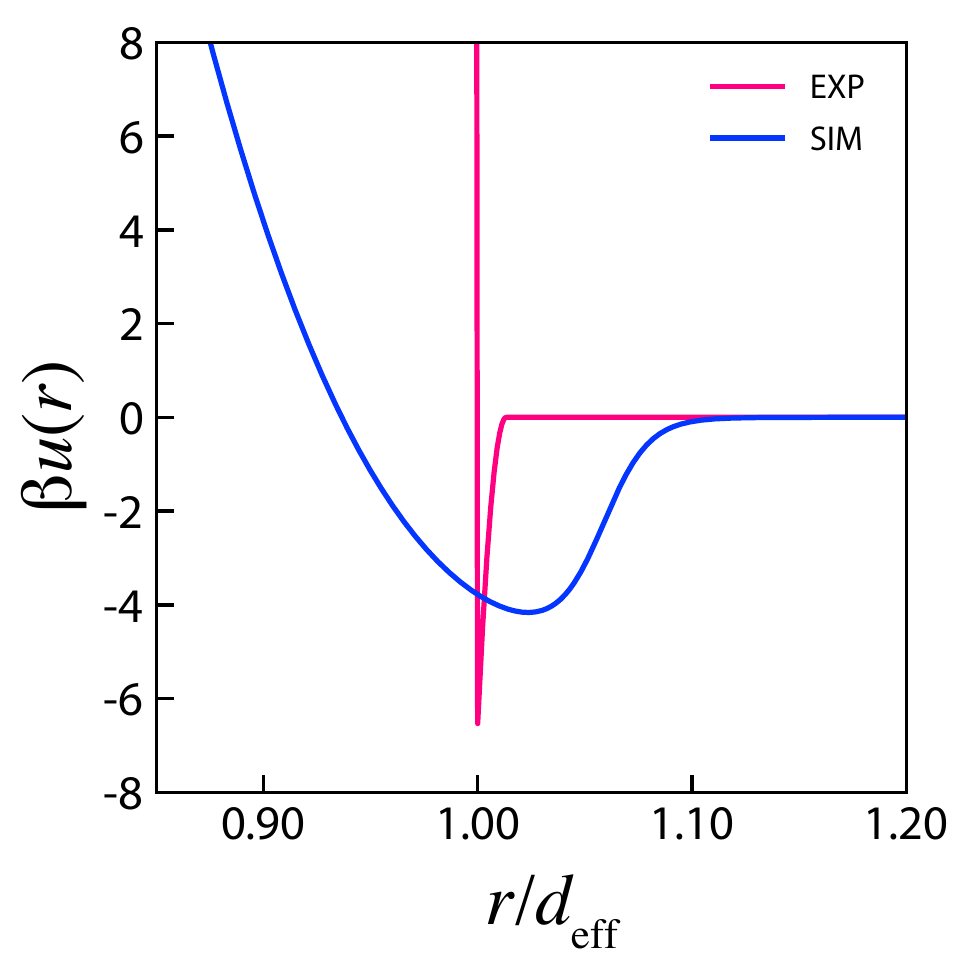}
\caption{The interaction potential used in the numerical simulations (Eq. \ref{eqsim}) and taken for the experiments (Eq. \ref{eqexp}) evaluated at criticality ($B_2^*=-3/2$). Here $\beta=1/k_BT$.}
\label{figInteractions}
\end{figure}

\subsection{Emulsion Preparation}

Colloidal polydimethylsiloxane (PDMS) emulsion droplets were synthesized following Elbers \textit{et al} \cite{elbers2015}. 
Sodium dodecylbenzenesulfonate (SDBS) surfactant (2 mM) and potassium chloride salt solution (20 mM) were added in order to stabilize PDMS emulsions and screen charges on droplet surfaces, respectively. The solvatochromic dye Nile Red was employed to fluorescently label PDMS emulsions. Glycerol is then added to obtain a refractive index matched emulsion with a weight ratio of water to glycerol around 51\% : 49\%. The droplets have mean diameter of $d=$3.2$\mu$m, which is determined from the first peak of the radial distribution function obtained from particle tracking. The Brownian time to diffuse a diameter 
\begin{equation}
\tau_B=\frac{\pi\eta d^3}{8k_BT}\approx19\mathrm{s}
\label{eqTaub}
\end{equation}
where $\eta$ is the solvent viscosity and $k_BT$ is the thermal energy.

\subsection{Colloid-Polymer Mixture Preparation}
The non-absorbing polymer utilised to induce depletion attraction is hydroxyethyl cellulose (Natrosol HEC 250 G Ashland--Aqualon) with molecular weight $3\times10^5$ g mol$^{-1}$. 
Colloid-polymer mixtures are prepared by adding stock solutions of HEC polymers (10 gl$^{-1}$) to concentrated emulsions with volume fractions around random close packing which we take to be $\phi_\mathrm{rcp}\approx0.64$.  All colloid-polymer mixture samples are observed under confocal microscope at about 16 $\tau_B$ after loading the sample cell. Our system is not density matched between droplets and solvent. In particular, colloidal systems, including depletion gels, are known to undergo batch settling (or, here, creaming) such that the local volume fraction in the bulk of the system is largely unaffected at short times  \cite{russel,royall2007prl,secchi2014,razali2017}. We ensure that we analyze data from the bulk of the sample where little change in volume fraction due to sedimentation is observed. Further sample details are listed in the Appendix.

\begin{figure}[hbt]
\centering
\includegraphics[width=85mm]{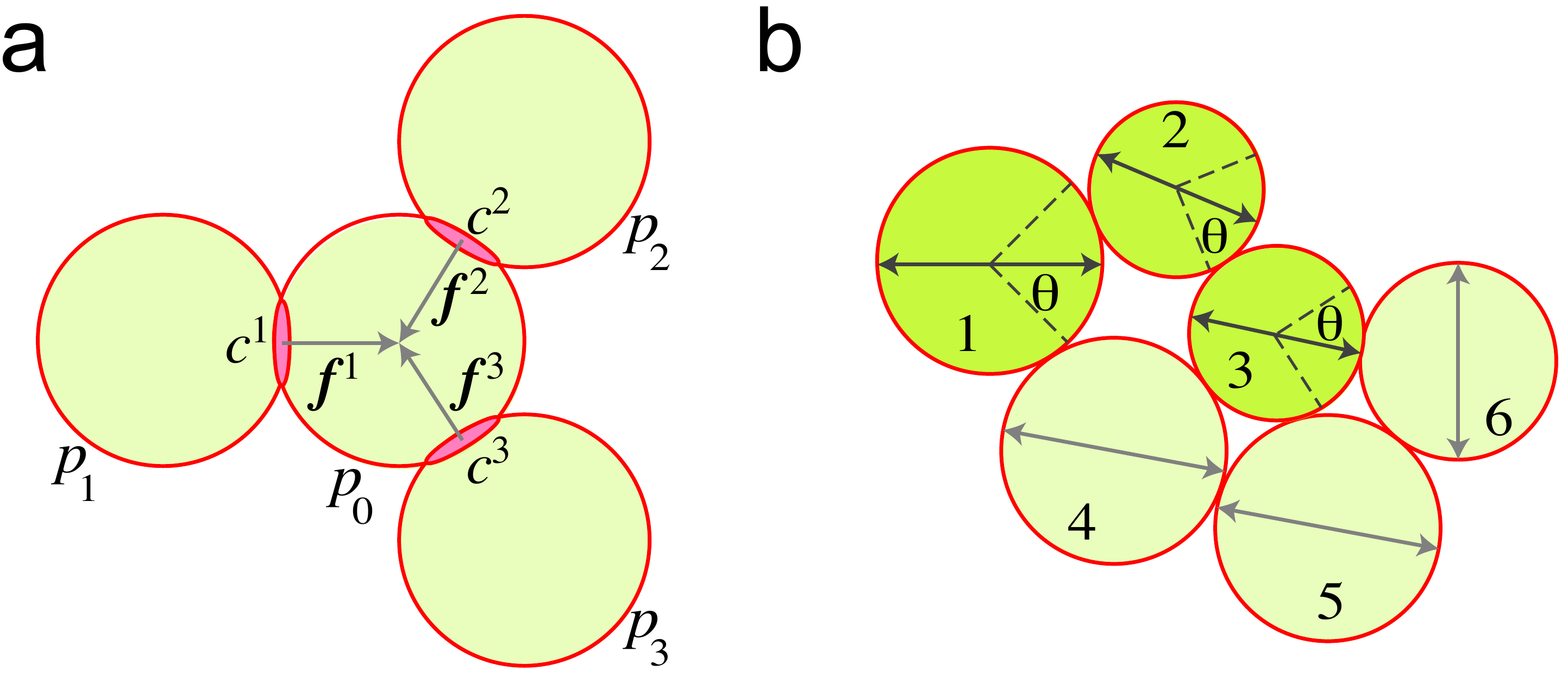}
\caption{\textbf{Local stress measure and force chains.}
(a) Schematic of contributions to the stress tensor from contacts $c_i$ and forces $\pmb{f}_i$ of neighboring particles $[p_1,p_2,p_3]$ of the particle of interest $p_0$.
(b) Schematic of our definition for force chains. Brightly colored particles [1,2,3] form part of a force chain. Particle 5 and 6 are not a member of the force chain because their centers lie more than $\theta$ away from the principal stress axis of particles 1 and 3 (indicated by the grey arrows). The centre of particle 6 \emph{does} lie within $\theta$ of the principal stress axis of particle 3, but its own principal stress axis lies at an angle greater than $\theta$ from that of particle 3, so it is not part of the force chain.} 
\label{figTense} 
\end{figure}

\subsection{Confocal imaging, particle and contact tracking}
\label{sectionTracking}
We used a Leica SP8 confocal microscope with a continuously tuneable white light laser. We use two--channel imaging with excitations 514 nm and 580 nm which correspond approximately to the absorption peaks of nile red in a non--aqueous and aqueous environment respectively. Nile red emits at differing frequencies in non--aqueous and aqueous environments with peaks of 545 and 645 nm respectively.  We exclude particles whose centres are closer than one diameter to the edge of the image to mitigate boundary effects.

To obtain information on the interdroplet contacts and forces we developed a method to mitigate the challenges resulting from the limited spatial resolution of the microscope. Previous work \cite{brujic2003,brujic2007} considered much larger droplets, but here we must contend with the challenge to optical microscopy posed by rather smaller colloidal droplets. The size of the contacts and in particular their separation from one another is comparable to the resolution of the microscope. We proceed by tracking the droplet coordinates \cite{leocmach2013} in the \emph{droplet} images [green channel in Fig. \ref{figContacts}(b)]. Since our system is reasonably monodisperse (polydispersity $\approx 8$ \%), we know that the contacts should be approximately equidistant between the centers of two neighboring droplets. The set of midpoints between neighboring droplets thus gives a trial set of candidate force--bearing contacts. Each of these is populated with a sphere, which we term a blob. From this we determine the magnitude of the force in the image by comparing with the number of pixels within this spherical volume and their intensity in the \emph{contact image} [red pixels in Fig. \ref{figContacts}(c)].

To obtain a measure of the force, we threshold the contact image. The contacts are identified on the basis of the number of pixels in the contact image that correspond to the ``blobs'' which are potential contacts. Our analysis gives a measure of the relative magnitude of the compressive force at contact points on each droplet. We compare our results to which are approximately matched to the experimental system. Further details of our analysis are given in the Appendix.

\subsection{Characterization of the droplet interactions}

The interaction between emulsion droplets is complex and depends on the local geometry \cite{lacasse1996}. Here we seek an 
estimate of the energy scales involved. Now the surface tension $\gamma=9.2$ mNm$^{-1}$ ~\cite{brujic2003}, which amounts to a energetic cost comparable to the thermal energy for a \emph{microscopic} change of surface area of the droplet. Therefore, in the case of our mesoscopic droplets, we expect deformations to be small. For such small deformations, we assume that two interacting droplets are deformed such that the surface in contact between them is a circle and determine the change in surface area with respect to two undeformed droplets of the same total volume. To leading order, the interaction energy

\begin{equation}
u_\mathrm{drop}(r)\approx\frac{\pi}{2}\gamma(d-r)^2
\label{eqdrop}
\end{equation}

\noindent
for $ r\le d$. Here $\beta=1/k_BT$. For our parameters, we expect that very small deformations around 0.1\% are sufficient to result in an interaction energy of many times that of the thermal energy (Fig. \ref{figInteractions}). Our droplets, therefore, approximate closely hard spheres \cite{royall2013myth}. Note that some other emulsion systems exhibit rather lower surface tension and therefore more deformation is found \cite{desmond2013}.

The polymer size is much smaller than that of the droplets, such that our system is towards the ``sticky sphere'' limit of short--ranged attraction strength. We presume that the effective attractions between the droplets are of the Asakura--Oosawa form, 
\begin{widetext}
\[
\beta 
u_\mathrm{AO}(r)=
\begin{cases}
\infty&$for $ r<d \\
u_\mathrm{drop}(r)+
\beta \frac{\pi (2R_g)^3 z_p^r}{6} \frac{(1+q)^3}{q^3}\left[1-\frac{3r}{2(1+q)d}+\frac{r^3}{2(1+q)^3d} \right] & $for $d \le r < d+(2R_g) \\
0 & $for $r \ge d+(2R_g) \\
\end{cases}
\label{eqAO}
\]
\end{widetext}
Here $q=2R_g/d$ is the polymer--colloid size ratio and $z_p^r$ is the polymer fugacity in a reservoir in thermodynamic equilibrium with the colloid--polymer mixture, which we assume to be equal to the polymer number density in the reservoir, as would be the case for ideal polymers. Here $R_g$ is the radius of gyration of the polymer. We neglect the contributions from electrostatics due to the Debye screening length which we estimate as 2nm which is much smaller than the range of the depletion attraction. Furthermore, using DLVO theory, we arrive at a contact potential due to electrostatic interactions between two droplets less than $k_BT$.  To estimate the interactions between the droplets, we assume that the attractive interaction remains for small compressions of the droplets $r<d$.

\begin{equation}
\beta u_\mathrm{exp}(r) = \begin{cases}
\beta u_\mathrm{drop}(r) + \beta u_\mathrm{AO}(d) &\text{if } r  \le d\\
\beta u_\mathrm{AO}(r)  &\text{if } d \le r.
\end{cases}.
\label{eqexp}
\end{equation}

\noindent
The interaction potential is plotted in Fig.  \ref{figInteractions}, where it is seen that the AO attraction is swiftly overwhelmed by the strong repulsion $u_\mathrm{drop}(r)$.

To proceed, we require the polymer radius of gyration and this we estimate from the gelation boundary. The phase diagram of our system is given in the Appendix in Fig. \ref{sFigPhase} in the polymer reservoir representation \cite{lekkerkerker1992}. We map our experimental values of the polymer concentration to the reservoir representation using Widom particle insertion \cite{lekkerkerker1992}. The polymer reservoir concentration corresponding to gelation $c_p^\mathrm{r,gel}$ is then 0.71$\pm 0.1$ $gl^{-1}$.  We express the polymer concentration as a ratio of this value. The polymer radius of gyration is then fixed by requiring that the reduced second Virial coefficient at the gelation boundary $B_2^*=-3/2$ ~\cite{noro2000}. While this holds for criticality, in fact for gels undergoing arrested spinodal decomposition (as is the case for colloid--polymer mixtures \cite{royall2021}), such short--ranged attractions lead to a very flat liquid--gas spinodal \cite{royall2018molphys,royall2018jcp}, such that the polymer concentration for gelation varies very little across a wide range of colloid volume fraction. In this way, we arrive a polymer $R_g=21.2$ nm and polymer--colloid size ratio $q=0.013$. This is close to the value quoted for HEC 250 G in the literature \cite{zhang2013}. The resulting effective droplet--droplet interaction potential is shown in Fig. \ref{figInteractions}. We are interested in the compressive forces between the droplets. We thus interpret these as $-d [\beta u_\mathrm{exp}(r)]/dr$ for $r  \le d$.

\subsection{Stress computation}
We now outline our method to obtain a measure of the local stress. Consider a reference particle $\pmb{p}_0$, for example with three neighbors $\pmb{p}_1$, $\pmb{p}_2$ and $\pmb{p}_3$ that touch through contacts $\pmb{c}_1$, $\pmb{c}_2$ and $\pmb{c}_3$, as shown in Fig. \ref{figTense}(a). The compressive force from particle $\pmb{p}_1$ is $\pmb{f}^1$, with magnitude $f^1$, which is determined from the size of contact $c^1$. Our single--particle stress measurement, $\pmb{\sigma}$ of a particle is calculated  by summing stress contributions from each neighbor on all element axes, indicated in Eq. \ref{eqStressTensor}. Our single particle stress measurement is a 3x3 matrix $\pmb{\sigma}$ whose elements are denoted by $\sigma_{ij}$, where $i,j$ label Cartesian components.  Similarly $\pmb{f}^c_i$ is the $i$th Cartesian component of the force on particle $c$:

\begin{equation}
\label{eqStressTensor}
\sigma_{ij} = \sum_{c=1}^{n^c}\pmb{f}_i^c\pmb{r}_j^c
\end{equation}

\noindent where $n^c$ is the number of contacts of $\pmb{p}_0$. Dividing this quantity by a suitable volume gives the Cauchy stress tensor, but here assigning the volume presents a challenge. As Fig. \ref{figContacts} shows, gels are heterogenous materials. Thus partitioning space according to a Voronoi decomposition leads to unphysically large separations. On the other hand, using the droplet volume does not fill space, as the volume fraction $\phi_c<1$. Here, we consider normalized quantities in reduced units where the mean particle diameter is set to unity. We shall therefore refer to $\pmb{\sigma}$ as a reduced stress tensor, noting that we apply it at the single--particle level. For each particle, we obtain 
$\pmb{\sigma}$ by analogy to the stress tensor, diagonalization generates three eigenvalues and eigenvectors, which represent principal stresses and principal directions respectively. After diagonalization, \paddyspeaks{the sum of all principal stresses}

\begin{equation}
\mathrm{tr}(\pmb{\sigma}) = \sigma_{xx} +\sigma_{yy} +\sigma_{zz}.
\end{equation}

\noindent
\paddyspeaks{Note that the quantity $-\mathrm{tr}(\pmb{\sigma})$ is analogous to the local pressure.}

\subsection{Force chain determination}
\label{sectionForceChainDetermination}

Here, to identify force chains, we consider a quasilinear assembly of at least three particles where stress is concentrated \cite{peters2005}. Based on this definition, a method was developed by Peters \textit{et al.} \cite{peters2005}, which we illustrate schematically in Fig. \ref{figTense}(b). If the minor principal stress of a particle is larger than the average compressive stress in the system, these particles are candidates for force chains. After selecting these particles with large stresses, we require that the particles with concentrated stresses should be quasilinear allowing only small amounts of rotation. Given a reference particle $1$, from its centre, we define a region that deviates of an angle of 
$\pm \theta$ 
from the direction of the stress $\pmb{\sigma}_A$. Here, we set $\theta=\pi/4$, we also require that the direction of symmetry on the second particle to be within $\theta$.

\begin{figure*}[hbt]
\centering
\includegraphics[width=180mm]{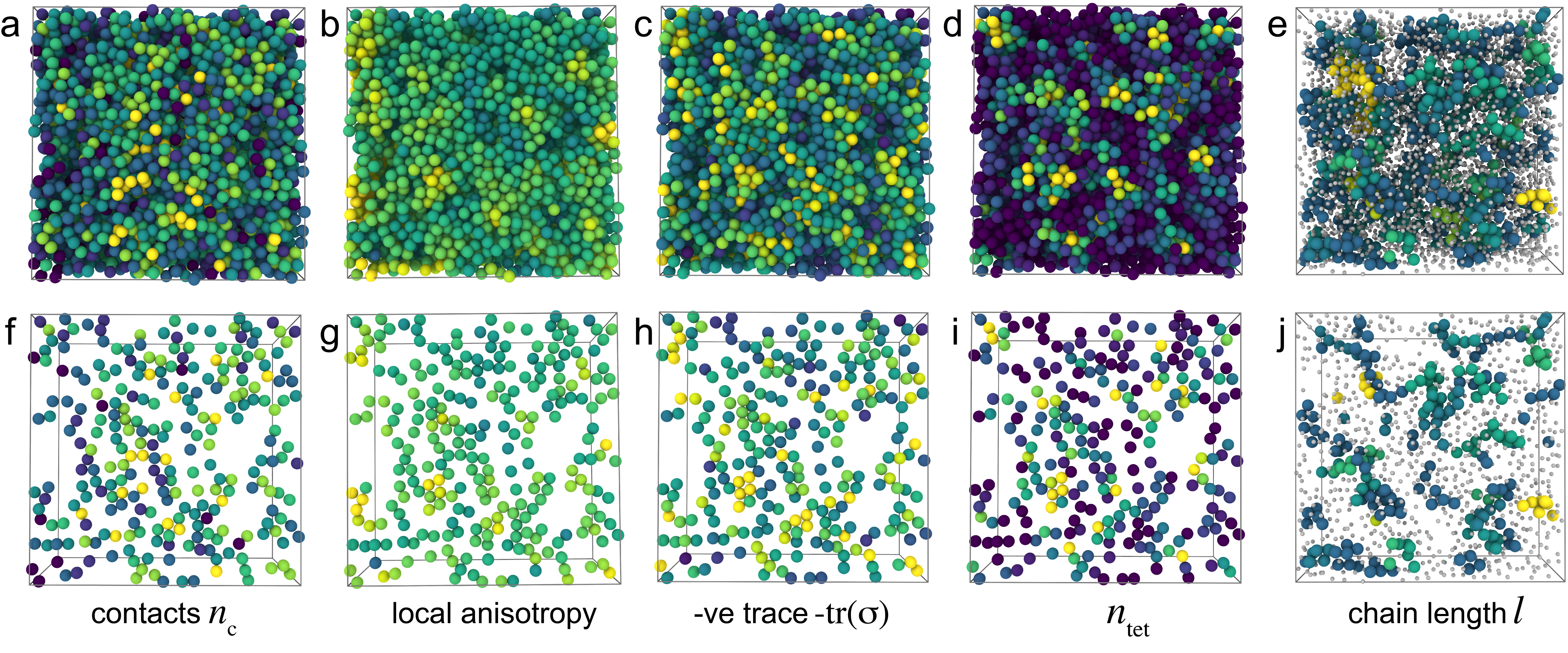}
\caption{\textbf{Renderings of quantities of interest} with each denoted underneath. 
(a) Number of contacts per particle. Maximum value (yellow) $n_c^\mathrm{max}=6$.
(b) Local anisotropy of the reduced stress tensor Maximum value (yellow) 0.5.
(c) Negative trace of the reduced stress tensor. Maximum value (yellow) $-\mathrm{tr}(\pmb{\sigma})^\mathrm{max}=2.0$.
(d) Number of tetrahedra a particle participates in. Maximum value (yellow) $n_\mathrm{tet}^\mathrm{max}=10$.
(e) Number of particles in force chains. Particles in force chains of $l=3$ or less are rendered small and grey. Maximum value (yellow) $l^\mathrm{max}=10$.
(f-j) Same data as (a-e), but a slice of thickness around one diameter $d$ is rendered instead, except (j) where a slice of thickness around $4d$ is rendered. All renderings are for experimental data with} $c_p^\mathrm{r}/c_p^\mathrm{r,gel}=1.5.$
\label{figPretty}
\end{figure*}

\subsection{Computer Simulation}
\label{sectionSimulation}

As noted above, to verify and calibrate our experimental data, we perform Brownian dynamics computer simulations. We use point particles interacting via a spherically-symmetric potential with Hertzian repulsive forces and a short-ranged attractive term, which we shift and truncate at a range $\rc$. The repulsive Hertzian contribution to the potential is

\begin{equation}
\beta u_{R}(r) = 
\beta A (1-r/d)^{5/2},
\end{equation}
while the attractive term is 
\begin{equation}
\beta u_{A}(r)=\dfrac{\beta\varepsilon}{2}\tanh\left(\dfrac{r/ d-1}{\delta} \right)
\end{equation}
so that the full potential is
\begin{equation}
\beta u_\mathrm{sim}(r) = \begin{cases}
\beta  u_R(r)+\beta u_A(r) & \\  
- \beta u_A(\rc) &\text{if $r< d$}\\
\beta u_A(r)-\beta u_A(\rc) &\text{if $d <r<\rc$}
\end{cases}
\label{eqsim}
\end{equation}

\noindent
The resulting $u_\mathrm{sim}(r)$ is plotted in Fig. \ref{figInteractions}.

To accurately model the short ranged, highly repulsive interaction between the droplets of the experiment, is highly challenging for conventional computer simulations. While novel methods have been developed for Monte Carlo simulations \cite{miller2003}, here we are interested in dynamical behavior. We therefore set $\beta A=1000, \delta = 0.02$ and $\rc=1.3 d$ which results in a very short range attraction with a soft core, see Fig. \ref{figInteractions}. The interaction strength $\varepsilon$ and the number density of the system $\rho$ characterize the state points. Using the Barker--Henderson effective hard sphere diameter
\begin{equation}
d_\mathrm{eff}=\int_0^{\infty}dr[1-e^{-\beta u_R(r)}], 
\end{equation}
we map number densities to effective volume fractions $\phi_\mathrm{eff}=\pi d_\mathrm{eff}^3\rho/6$. We presume that the effective volume fraction corresponds to the absolute droplet volume fraction in the experiments.

Simulations are performed in the isothermal-isochoric ensemble (NVT) solving the Langevin dynamics

\begin{equation}
m\dot{v}_i=-\nabla_i u-\gamma{v}_i+\sqrt{2\gamma/\beta}{\xi}_i
\end{equation}
for particles of equal mass $m$ in the presence of a zero-mean, unit-variance random force $\xi$. To this purpose, we employ a suitably modified version of the LAMMPS molecular dynamics package.

We use a velocity--Verlet integrator with timestep $dt=0.001\tau_0$ with $\tau_0=\sqrt{m\beta}d$. Fluctuations of the temperature are allowed to damp on a relatively short timescale of $\tau_d=100dt= 0.1\sqrt{m\beta}d$, a setup for which results are similar to the overdamped limit \cite{razali2017}.  This timescale also sets the Brownian time $\tau_B=\gamma d^2/24k_BT=\tau_0^2/24\tau_d$, which allows us to compare the numerical results with the experiments via Eq. \ref{eqTaub}.

As non--equilibrium systems, gels coarsen over time \cite{royall2021,manley2005}. Now there is a significant difference in this coarsening process between experiments and Brownian dynamics simulations, due to hydrodynamic interactions in the former which are not fully accounted for in the latter \cite{furukawa2010,royall2015prl,varga2016,degraaf2019}. Therefore, even if a precise matching of timescales were to be carried out, in fact one would still expect considerable differences between experiment and simulation, as has been found previously \cite{royall2015prl}. For our purposes, then, we select a point in the time--evolution of the simulations of $12\tau_B$, in which we find that a number of time--dependent properties are comparable to those in the experiments (see section \ref{sectionResults}). We set the volume fraction $\phi_\mathrm{eff}=0.2$, and to compare the interaction strength with the experiments, we scale by the value corresponding to gelation $\varepsilon^\mathrm{gel}$. Following the experiments, we set $\varepsilon^\mathrm{gel}$ as that at which the reduced second virial coefficient $B_2^*=-3/2$ \cite{noro2000,royall2018jcp}. Given that the interaction potential in the simulations is also rather softer than that of the experiments, we regard comparison between our two approaches to be semi--quantitatively, rather than the simulations being an accurate reproduction of the experiments.

\section{Results}
\label{sectionResults}

We organize our results section as follows. We begin by discussing the structural properties, the number of neighbors and the contacts. We then move on to consider the forces between droplets inferred from the contacts, leading to quantities such as the reduced stress tensor. We then consider force chains. Throughout, we compare our experimental results with those from computer simulation.

\begin{figure*}[hbt]
\centering
\includegraphics[width=185mm]{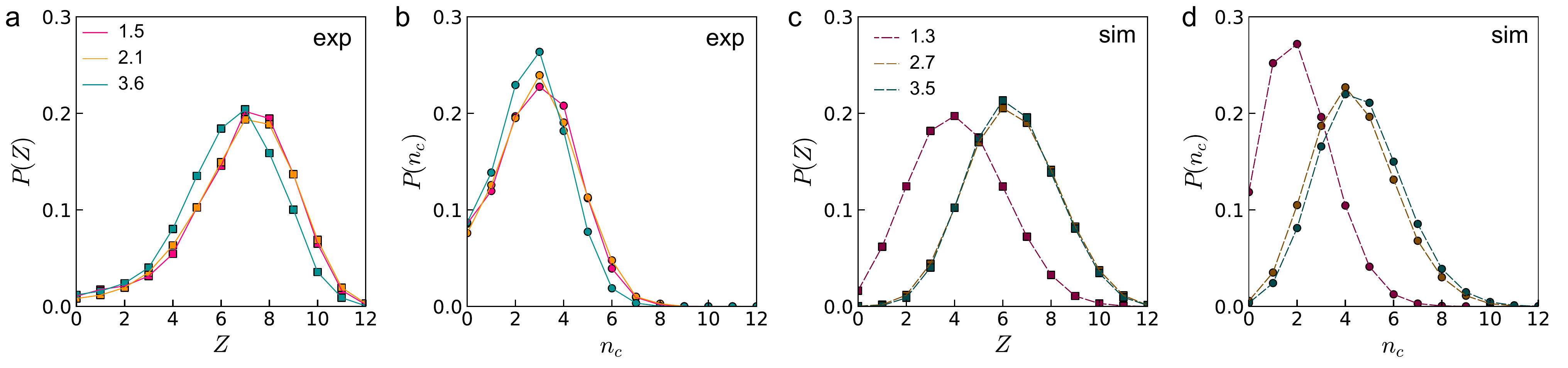}
\caption{\textbf{Local coordination and contacts.} 
Distribution of of number of nearest neighbors within the first coordination shell $Z$ in experiments (a) and simulations (c). 
Number of contacts $n_c$ in experiments (b) and simulations (d). 
The experimental data is shown at various $c_p/c_p^\mathrm{r,gel}$ and simulation at $\varepsilon/\varepsilon^\mathrm{gel}$.
The line colors in (a,b) and (c,d) are common and are used throughout the remainder of this article.
}
\label{figNNContact}
\end{figure*}

\subsection{Neighbors and Contacts}
\label{sectionNeighbors}

We consider, schematically, the imaging methodology and method for interparticle force extraction in Fig. \ref{figContacts}(a). Representative data of each fluorescent channel is shown in Figs. \ref{figContacts}(b, c), and their combination in \ref{figContacts}(d). We render the droplets actual size and, following the identification of contact analysis outlined in section \ref{sectionTracking} and described in more detail in the Appendix, the contacts as pink sticks in Fig. \ref{figContacts}(e). This constitutes our basic data. Having demonstrated the principles of our method, we consider quantities of interest.

We proceed to show renderings of properties of particular interest in Fig. \ref{figPretty} for a polymer concentration of $c_p/c_p^\mathrm{r,gel}=1.5$. Other gel state points appear similar. We show the number of contacts for each particle $n_c$, which appears to be rather heterogeneous throughout the system. Before considering the other quantities, we move to a quantitative discussion of the coordination and the number of contacts around Fig. \ref{figNNContact}.

In Fig. \ref{figNNContact}(a), we show the distribution of the number of neighbours, $Z$, which are defined as being closer than $1.2d$, which is close to the first minimum of the radial distribution function $g(r)$. The number of neighbors requires knowledge of only the droplet coordinates and thus comparison can be made to other work with particle--resolved studies, and indeed similar behavior can be found, e.g. in Fig. 2 of ref. \cite{ohtsuka2008}. The number of contacts $n_c$ for the same data points is shown in Fig. \ref{figNNContact}(b). This has a smaller value to the number of neighbors. The distributions of neighbors and contacts in our simulations show very similar behavior, as shown in Fig.  \ref{figNNContact}(c,d). In the simulations, the $\varepsilon/\varepsilon^\mathrm{gel}=1.3$ state point has fewer neighbors and contacts than the others we have shown. However, it is worth nothing that this is rather closer to the gelation boundary than the others (the next closest being the experimental state point at  $c_p/c_p^\mathrm{r,gel}=1.5$), which could account for the difference.

\subsection{Forces}
\label{sectionForces}

\begin{figure*}
\centering
\includegraphics[width=170mm]{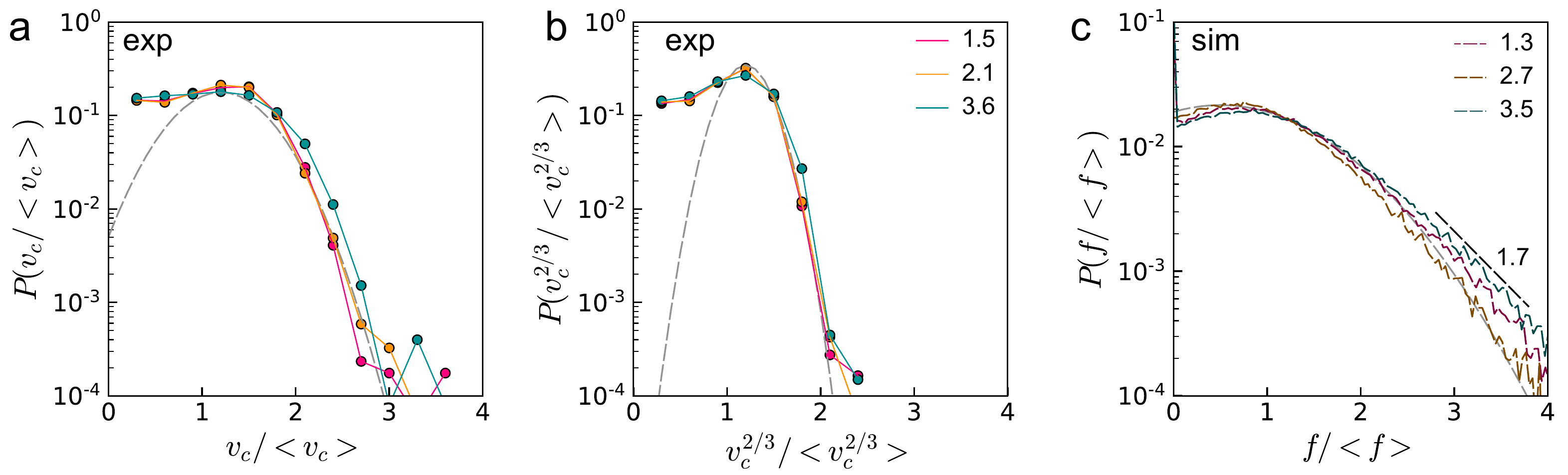}
\caption{\textbf{Distribution of forces.} Comparison between experiments and simulations for the distributions of contact volumes $v_c$ (a) and $v_c^{2/3}$ (b) for the experiments and repulsive forces for the simulations (c) Here the angle brackets denote the mean of the distribution. The experimental data is shown at various $c_p/c_p^\mathrm{r,gel}$ and simulation at $\varepsilon/\varepsilon^\mathrm{gel}$. Dashed grey lines are Gaussian plots, and the black dashed line in (c) $\sim\exp(-1.7\langle f \rangle)$.}
\label{figForces}
\end{figure*}

We can estimate the relative compressive forces on each droplet. At the level of our analysis, we determine the number of pixels in the contact image within a ``blob'' (see Appendix and Fig. \ref{sFigContactTrackingSchematicWeighted}). We take the sum of these contact pixels as a measure of the contact volume $v_c$, which we plot in Fig. \ref{figForces}. For granular systems, the force has been identified with the contact area \cite{brujic2003,jorjadze2013}, which should scale as $v_c^{2/3}$. We therefore plot the distribution of $v_c^{2/3}$ in Fig.  \ref{figForces}(b). This is much sharper than the distribution of volumes. In our simulations, we have direct access to the compressive forces, and these we plot in Fig. \ref{figForces}(c). The distribution from the simulations is rather broader. Indeed, except for the smallest forces, the experimental data is roughly compatible with a Gaussian distribution. However there is some evidence in the simulations for an exponential decay [black line in Fig. \ref{figForces}(c)]. As noted above, although the time--evolution in the experiments and simulations differs, the structural quantities in Fig. \ref{figNNContact} are rather similar across the two systems. Therefore, it is possible that the difference in the interaction potential between the two (Fig. \ref{figInteractions}) may underlie the difference between the force distributions that we obtain. Since the measured contact volumes also depend on the particle dynamics and the imaging process, it is likely that they do not respond to very fast force fluctuations.  This would mean that the force inferred from the contact size corresponds to a time-averaged version of the interparticle force. The time-averaging will act to suppress force fluctuations.  This suppression is absent from simulations, where the (instantaneous) microscopic force is measured directly.

A further possibility is that the contacts in the experiments are imaged over a certain period, which in fact corresponds to several $\tau_B$ per particle (imaged in the $z$ direction, $xy$ planes are acquired rather more quickly). Therefore, there is some averaging of the experimental data, which is absent from that shown in the simulations which correspond to a single snapshot.

\begin{figure}[hbt]
\centering
\includegraphics[width=85mm]{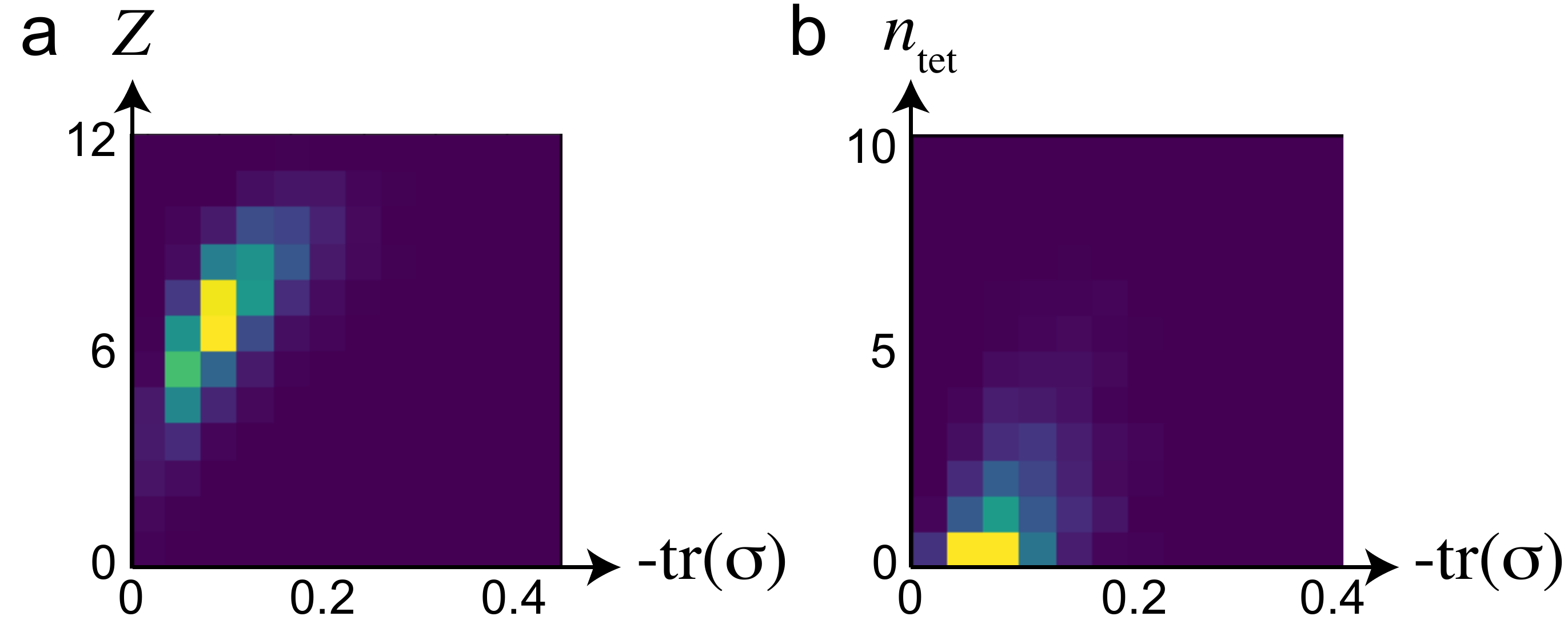}
\caption{\textbf{Correlations between some quantities of interest.} 
(a) Heat map of the number of neighbors $Z$ and negative trace $-\mathrm{tr}(\pmb{\sigma})$.
(b) Heat map of $-\mathrm{tr}(\pmb{\sigma})$ and the number of particles participating in tetrahedra $n_\mathrm{tet}$.
Data are shown for an experimental system with $c_p^\mathrm{r}/c_p^\mathrm{r,gel}=1.5.$} 
\label{figCorrelations}
\end{figure}

Identification of the forces associated with each contact allows us to investigate the reduced stress tensor $\pmb{\sigma}$, Eq. \ref{eqStressTensor}. In Fig. \ref{figPretty}(b), we show the \emph{local anisotropy}, which is the difference between largest and smallest eigenvalues of $\pmb{\sigma}$. Although it may appear from visual inspection that this quantity has some spatial correlation, we have investigated such correlations and find that these are indistinguishable from the (short--ranged) density correlations expressed via the radial distribution function. We then plot the negative trace of the reduced stress tensor $-\mathrm{tr}(\pmb{\sigma})$ which is analogous to the local pressure in Fig. \ref{figPretty}(c). Like the number of contacts [Fig. \ref{figPretty}(a)], this is rather heterogeneous. The trace is correlated with the number of neighbors, with higher pressure corresponding to a larger number neighbors [Fig. \ref{figCorrelations}(a)]. Here the Pearson correlation coefficient is 0.709.

Colloidal gels have been subjected to structural analysis, in particular clusters which are local energy minima have been identified with rigidity \cite{royall2008,royall2021}. Now the local structure changes over time, leading to larger and more complex local structure \cite{royall2012,royall2018jcp}, and at early stages like the gels of interest here the dominant local structure is the tetrahedron \cite{royall2015prl}. It is possible to classify the particles according to the \emph{number} of local structures in which they participate, which can reveal the degree of local ordering \cite{dunleavy2015,hallett2020}. Here therefore, we count the number of tetrahedra in which each particle participates, as shown by the rendering in Fig. \ref{figPretty}(d). Visual inspection suggests that there is some correlation between the number of tetrahedra the particles participate in, and the trace $-\mathrm{tr}(\pmb{\sigma})$  [Fig. \ref{figPretty}(c).]. This is indeed the case [Fig. \ref{figCorrelations}(b)] with the correlation coefficient being 0.455.

\subsection{Force Chains}
\label{sectionChains}

\begin{figure}
\centering
\includegraphics[width=60mm]{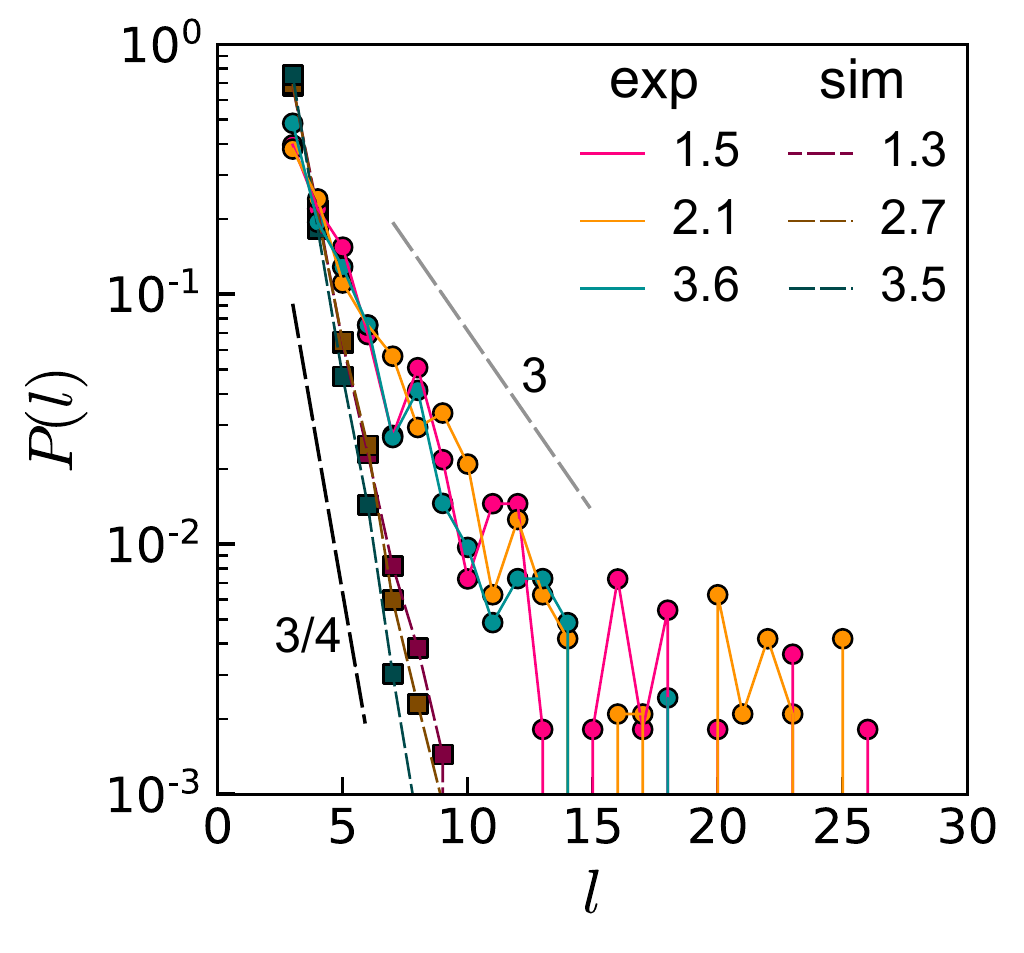}
\caption{\textbf{Force chain length.}
Distribution of force chain lengths for experiment and simulation. State points are indicated for experiment as $c_p/c_p^\mathrm{gel}$ and simulation as $\varepsilon/\varepsilon^\mathrm{gel}$. Dashed grey and black lines exponentials with decay lengths indicated.}
\label{figChainDist} 
\end{figure}

We implement the measurement of force chains outlined in section \ref{sectionForceChainDetermination}. In this way, we obtain the distribution of force chain lengths $P(l)$ in our system. We note that there is no reason \emph{a priori} to expect that these would span the system, as is the case for granular materials in compression \cite{majmudar2005}. In fact the majority of particles are found in force chains of a single particle. Longer force chains are rendered in Fig. \ref{figPretty}(e). When we plot the distribution of force chain lengths $l$ in Fig. \ref{figChainDist}, we find that in both simulation and experiment, that the effect of interaction strength is weak. The force chains in experiment are rather longer. Our data are compatible with an exponential distribution, with a decay length of $3$ and $3/4$ in experiment and simulation respectively.

Note that here we may cut some force chains at the image boundaries. Although we neglect contributions closer than a diameter $d$ to the boundary, it is hard to remove possibly boundary effects from the force chain distribution for images or the size that we acquire here. However, we may observe that in Fig. \ref{figPretty}(e), the force chains are rather smaller than the imaging volume and thus we expect any boundary effects to be reasonably small, and in any case, these will tend to reduce the apparent chain length, so such boundary effects are unlikely to be the cause of the difference between the experiments and simulation that we see. Given that hydrodynamic interactions are associated with more linear structures \cite{furukawa2010,royall2015prl,degraaf2019}, it is tempting to suppose that these are part of the reason for the longer chains that we find in the experiments.

\section{Discussion and Conclusions}
\label{sectionDiscussion}

We have characterized the interparticle contacts in a colloidal gel of emulsion droplets. We have further investigated compressive forces between droplets related to these contacts, and have semi--quantitatively benchmarked our results against computer simulation. We have fewer contacts and particles with large numbers of contacts are not strongly correlated in space.

Turning towards the forces, these we infer from the number of pixels in the blobs in the contact image which measures the spatial distribution of solvatochromic dye. The change in droplet surface area due to deformation of the mesoscopic emulsion droplets incurs a high energetic cost, as the surface tension is of the order of the thermal energy for a microscopic (molecular) change in surface area. Under the depletion forces due to the polymer, we therefore expect very weak deformation of the droplets. We believe that the contact volume inferred from the images of the solvatochromic dye is larger than the true contact area. Further investigations in this direction are clearly desirable, perhaps using systems with lower surface tension whose droplets would be deformed rather more \cite{desmond2013}. Nevertheless, the normalized force distributions that we obtain are comparable to our simulations. The somewhat broader distribution in the simulations might be related to the softer interaction potential that we have used. This width could be (somewhat) narrowed towards that assumed for the experiments to investigate if this is the cause of the difference.

We have obtained a measure for the local pressure from the reduced stress. Like the number of contacts, this is not strongly correlated in space. However, it is quite well correlated with the number of neighbors and also with the local structure, as expressed by the number of tetrahedra that a droplet participates in.

The force chains that we find in this thermal system are rather shorter than those encountered in granular systems with repulsive interactions \cite{majmudar2005}. Again, we encounter similar behavior in simulation, although the force chains are somewhat  longer in our experiments, which may be related to hydrodynamic interactions in the latter which are largely neglected in the former. The effect of HI would thus be interesting to probe in the future. While granular systems with attractive interactions have been investigated, there the focus lay more towards the contacts \cite{jorjadze2011}. Given the much higher volume fraction investigated in that work, direct comparison is hard, not to mention the differences between the thermal and athermal nature of the systems. It would nevertheless be most attractive to explore the force chain distribution in attractive jammed materials, such as granular gels \cite{li2014}. Granular systems with repulsive interactions are by their nature found at high volume fraction, and force chains typically percolate to form force networks. Nevertheless, there is some evidence for an exponential distribution in community sizes \cite{bassett2015} in force networks, the same scaling as we find the much shorter linear chains.

Thus we present a colloidal version of a model system for characterizing contacts and interdroplet forces. By considering perturbation such as shear, this system may be used to obtain a knowledge of local stress that may prove useful in understanding failure in soft solids such as colloidal glasses and gels.

\textbf{Acknowledgements}

We thank Paul Bartlett, Jasna Bruji\'{c} and Jens Eggers for helpful discussions and Yushi Yang his valiant assistance with the TCC analysis. CPR acknowledges the Royal Society for support, JD, FT and CPR acknowledge European Research Council (ERC Consolidator Grant NANOPRS for support, project number 617266). JD acknowledges Bayer AG for support. \paddyspeaks{RLJ and CPR acknowledge EPSRC for support via EP/T031247/1.}
EPSRC grant code EP/ H022333/1 is acknowledged for provision of the confocal microscope used in this work.


%

\appendix*

\section{Details of the Acquisition and Analysis of the Experimental Data}

To image the system with confocal microscopy, we employed two excitation lasers with wavelengths of 514 nm and 580 nm and two HyD hybrid detectors with detection wavelengths of 520--575 nm and 585 --640 nm. These two lasers and detectors are applied to detect fluorescent signals from bulk of the PDMS oil droplets and the contacts between the droplets, respectively. We refer to the images generated by these two channels as the \emph{droplet image} and \emph{contact image} respectively. We equalized the image histograms as a function of depth 
to compensate for any attenuation due to imperfect refractive index matching between emulsion droplets and solvent. To reduce the noise of the captured images, we applied line averaging of 32 to each frame and deconvolved the images with the Huygens software. Droplet centres were detected in the droplet image using colloids tracking package \cite{leocmach2013}.

\begin{figure}[htbp!] 
\centering    
\includegraphics[width=65 mm]{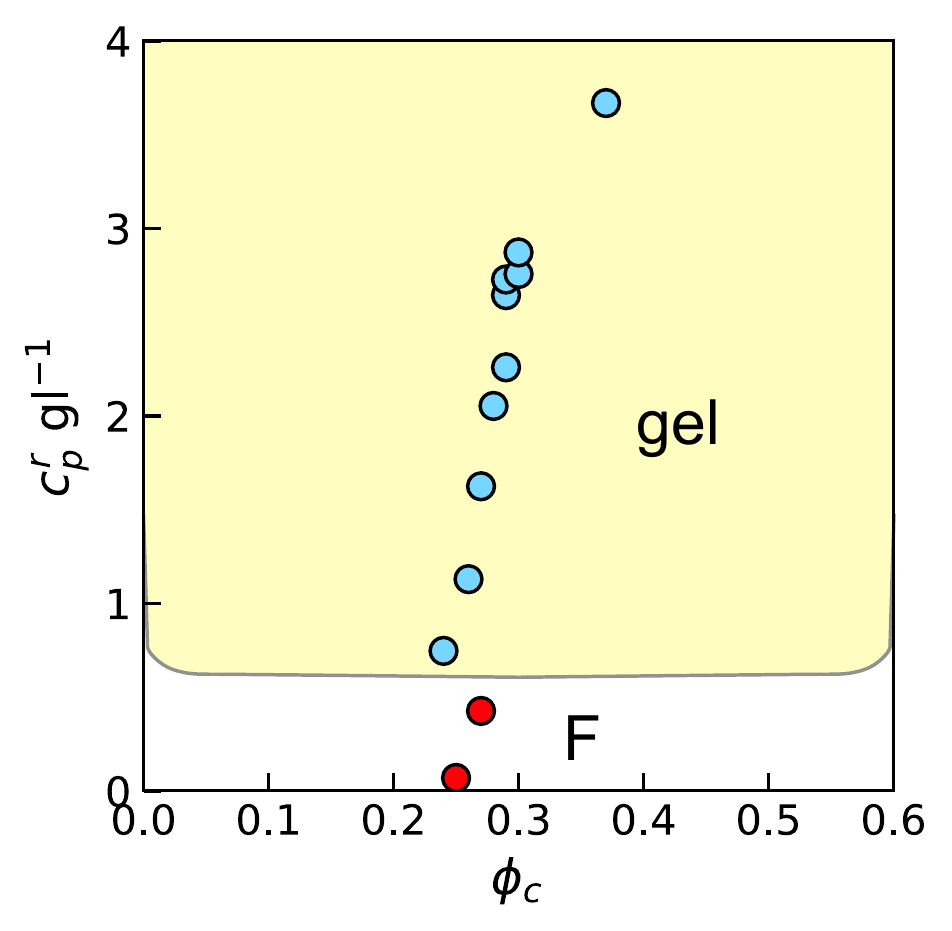}
\caption{
Experimental phase diagram in the colloid volume fraction -- polymer reservoir concentration plane.}
\label{sFigPhase}
\end{figure}

\begin{figure}[htbp!] 
\centering    
\includegraphics[width=85 mm]{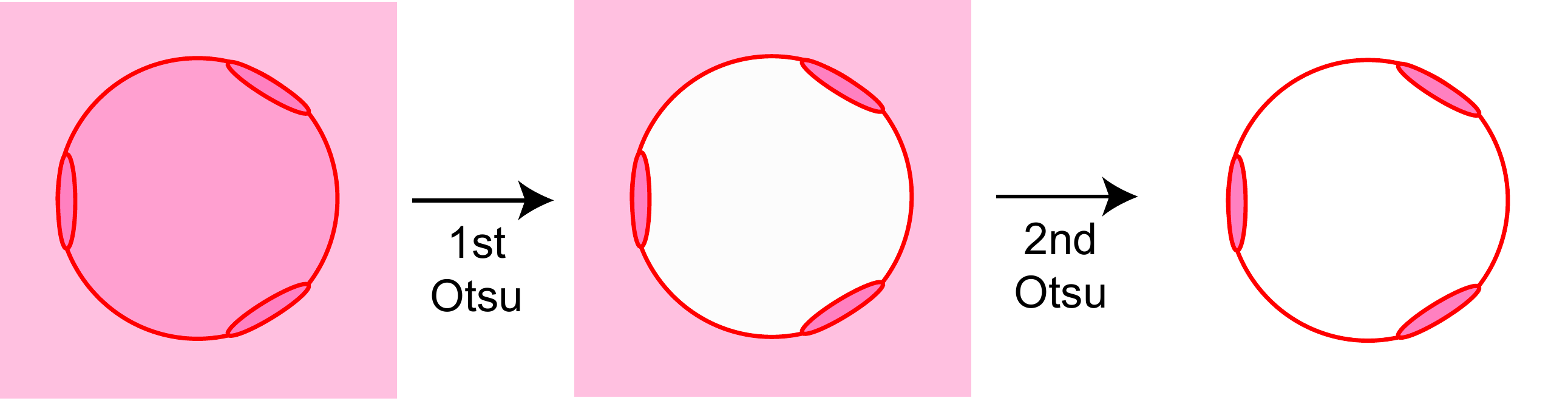}
\caption{
Schematic of the image processing for the enhancement of contact images by the use of two Otsu thresholds.}
\label{sFigContactTrackingSchematicOtsu}
\end{figure}

\subsection*{Tracking of interparticle contacts}

Here the smaller lengthscale with respect to previous work with much larger droplets \cite{brujic2003} necessitates a method to segregate connected contacts, determine centres and sizes of contacts. Having obtained the droplet centres, we proceed by processing the contact images. We use two Otsu thresholds (which is a threshold based on weighted variances of intensities of pixels corresponding to features and background \cite{otsu1979}). As schematically shown in Fig. \ref{sFigContactTrackingSchematicOtsu}, we apply two Otsu thresholds to the contact images. The first distinguishes droplets (with contacts) from the solvent background. The second separates contacts (foreground) from bulk droplets (background).

\textit{Edge enhancement ---} To remove any contacts erroneously identified due to residual intensity in the interior of the droplets, we use a Sobel filter to enhance the droplet edges. However applying the Sobel filter directly to the droplet image means that the edges of each particle are not always well defined because some droplets are in contact with one another. Therefore instead we generate an image from the particle coordinates we have determined and apply the Sobel filter to each particle.

\begin{figure*}[htbp!] 
\centering    
\includegraphics[width=170 mm]{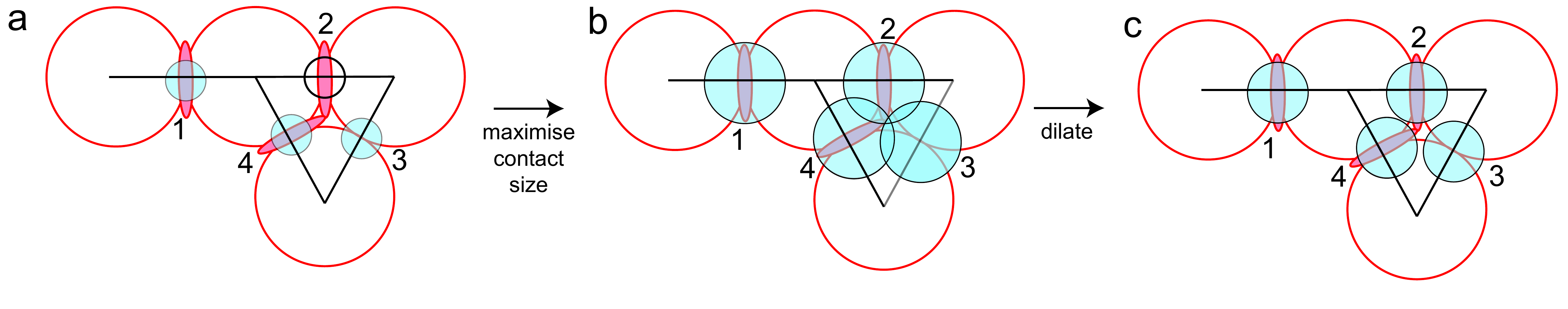}
\caption{Identifying contacts.
(a) Positioning of ``blobs'' at potential contact points, defined as the weighted middle points between droplets. Weighted middle points of the black lines connecting 
droplet centers are then potential contacts, as identfied with blue spheres and numbered.
(b) Blobs are expanded to include all contact area. 
(c) Blobs are dilated to remove overlaps. The combination of (b) and (c) then contains the information of contacts labels and, by reference to the contact image, contact volumes.}
\label{sFigContactTrackingSchematicWeighted}
\end{figure*}

\textit{Weighted middle points between particles --- } 
After thresholding images such as Fig. \ref{figContacts}(c), we find that the contacts are frequently merged. In order to separate such connected contacts, our strategy is to add a spatial boundary to each contact. The first step is to find the weighted middle points between a reference particle and its neighbors, which are possible locations of contact centers. To determine the weighted middle points $\mathbf{m}_{ij}$ between two neighboring droplets $i,j$, first $\mathbf{m}_{ij}$ needs to be located on the line connecting the centers of droplets $i$ and $j$ and the distances between $\mathbf{m}_{ij}$ and the two neighboring particles $b_i$ and $b_j$ are proportional to particle radii $a_{ij}$, \textit{i.e.} $b_i/b_j = a_i/a_j$. Therefore a binary mask of the same size of the contact image is built, where the positions of weighted middle points $\mathbf{m}_{ij}$ have a value of 1 while the rest of the mask is 0.

\textit{Positioning blobs on middle points --- } 
Based on centers of middle points, spherical blobs were created by dilating a binary kernel in three dimensions. The blobs were constructed as large as possible but without overlapping with each other. The purpose of building blobs is to contain true contacts as much as possible and build an upper boundary for the contacts to separate them from each other if they are overlapping after the thresholding. Because blobs are created in between neighboring particles (which are not necessarily in contact), so the number of blobs generated is greater than the number of true contacts.

After the initial placement of blobs [Fig. \ref{sFigContactTrackingSchematicWeighted}(a)], some are connected when we try to maximize their size as shown in [Fig. \ref{sFigContactTrackingSchematicWeighted}(b)]. By looking at the distribution of blob volumes, it is clear that connected blobs have noticeably larger volumes than isolated blobs, the binary mask with all blobs was separated into two masks: a well separated blob mask [Fig. \ref{sFigContactTrackingSchematicWeighted}(b), blob ``1''] and a connected blob mask [Fig. \ref{sFigContactTrackingSchematicWeighted}(b), blob ``2,3,4''], In the mask with connected blobs, we eroded the mask in order to separate these blobs  [Fig. \ref{sFigContactTrackingSchematicWeighted}(c)]. Next, an eroded mask [Fig. \ref{sFigContactTrackingSchematicWeighted}(c)] and non-connected blob mask [Fig. \ref{sFigContactTrackingSchematicWeighted}(b), blob ``1''] were combined into a final binary mask. This mask effectively sets 
bounds for contacts and can be used to segregate connected contacts. At this point, we have identified the contacts. However, we now seek to to determine their size, from which we can infer the force related to each contact.

\textit{Centres and sizes of contacts --- }
Three masks are generated in order to correctly detect the positions and sizes of contacts. The first mask, [Fig. \ref{sFigContactTrackingSchematicWeighted}(a)], is the binary mask of spheres that are placed between particles. This mask segregates some contacts that are connected after the Otsu thresholding of the contact image. It is possible that some pixels which are located in the middle of particles remain after the thresholding. Therefore a second mask which contains edges of all particles is desired, in order to set constraints to contact positions. This means contacts can only be located at edges of particles but not inside particles. The third mask is the thresholded contact image, which is obtained by applying the Otsu threshold to the contact image [Fig. \ref{sFigContactTrackingSchematicOtsu}]. By convolving these three masks, the remaining pixels are the contacts between droplets. Each contact is then labelled with an index, and by counting the number of pixels in each contact then gives the volume of the contact. The contact centre is determined by finding the geometrical centre or maximum intensity pixel in the contact.

\textit{Allocation of contacts to particles --- }
After particle and contact tracking, both coordinates and sizes are obtained. The coordinates of particles and contacts are $ \pmb{p}_i $ and $ \pmb{c}_j $ respectively. The distances between each particle and contact are computed, and stored in a $ i \times j $ 2d matrix $\pmb{s}_{ij}$.
\begin{equation}\label{eqParticleDistance}
\pmb{s}_{ij} = \lbrace i \in N_p, j \in N_c \mid \mid  \pmb{p}_i - \pmb{c}_j \mid   \rbrace
\end{equation}
where $  N_p $ and $ N_c $ are the number of particles and contacts respectively. For a contact $\pmb{c}_j$, the closest two particles $\pmb{p}_a$ and $\pmb{p}_b$ are detected by searching for the first two minimum values $ s_{aj}$ and $ s_{bj}$ in $ \pmb{s}_{ij}$. These two particles are then in contact through 
$\pmb{c}_j$. For each contact, we find two neighbor particles, in turn we can determine neighbour contacts for each particle, and this gives the number of contacts $n_c$.

\end{document}